\documentclass[conference]{IEEEtran}

\usepackage{cite}
\usepackage{amsmath,amssymb,amsfonts}
\usepackage{graphicx}
\usepackage{textcomp}
\usepackage{xcolor}
\usepackage{comment}
\usepackage{tikz}
\usepackage{orcidlink}
\usepackage{booktabs}
\usepackage{tabularx}
\usepackage{bm}
\usepackage{lipsum} 
\usepackage{caption} 
\usepackage{subcaption}
\usepackage{float}
\usepackage[noend]{algpseudocode}
\usepackage{algorithm}

\ifCLASSINFOpdf
\else
\fi

\def\BibTeX{{\rm B\kern-.05em{\sc i\kern-.025em b}\kern-.08em
    T\kern-.1667em\lower.7ex\hbox{E}\kern-.125emX}}
\begin{document}

\title{ Enhancing AI Transparency: XRL-Based Resource Management and RAN Slicing for 6G ORAN Architecture
}
\author{
    \IEEEauthorblockN{
         Suvidha Mhatre\orcidlink{0000-0003-2601-2774}\textsuperscript{*},
         Ferran Adelantado\orcidlink{0000-0002-9696-1169}\textsuperscript{†},
         Kostas Ramantas\orcidlink{0000-0002-1304-784X}\textsuperscript{‡},
        Christos Verikoukis\orcidlink{0000-0001-8774-1052}\textsuperscript{§}
    }
    \IEEEauthorblockA{
        \textsuperscript{*}\textit{Universitat Politecnica de Catalunya, Barcelona, Spain}
        \textsuperscript{†}\textit{Universitat Oberta de Catalunya, Barcelona, Spain}\\
        \textsuperscript{‡}\textit{Iquadrat Informatica S.L., Barcelona, Spain}
        \textsuperscript{§}\textit{University of Patras and ISI Athena, Greece} \\
    }
}

\maketitle
\begin{abstract}
This research introduces an advanced Explainable Artificial Intelligence (XAI) framework designed to elucidate the decision-making processes of Deep Reinforcement Learning (DRL) agents in ORAN architectures. By offering network-oriented explanations, the proposed scheme addresses the critical challenge of understanding and optimizing the control actions of DRL agents for resource management and allocation. Traditional methods, both model-agnostic and model-specific approaches, fail to address the unique challenges presented by XAI in the dynamic and complex environment of RAN slicing. This paper transcends these limitations by incorporating intent-based action steering, allowing for precise embedding and configuration across various operational timescales. This is particularly evident in its integration with xAPP and rAPP sitting at near-real-time and non-real-time RIC, respectively, enhancing the system's adaptability and performance. Our findings demonstrate the framework's significant impact on improving Key Performance Indicator (KPI)-based rewards, facilitated by the ability to make informed multimodal decisions involving multiple control parameters by a DRL agent. Thus, our work marks a significant step forward in the practical application and effectiveness of XAI in optimizing ORAN resource management strategies.

\end{abstract}


\section{Introduction}
Future 6th generation (6G) network and envisioned services demand diverse and high key performance indicators (KPIs) under different use cases defined by 3rd generation partnership project (3GPP). These use cases—such as enhanced mobile broadband (eMBB), ultra-reliable low latency communication (URLLC), and massive machine-type communication (mMTC)—outline distinct requirements for the efficient operation of future networks. To meet these expectations, it is essential for 6G networks to incorporate dynamic and optimal management of resources, particularly in Radio Access Network (RAN) environments. There are numerous challenges in radio resource management (RRM) that can benefit from the application of machine learning (ML) techniques, particularly deep reinforcement learning (DRL). One promising approach within DRL is the use of deep Q-networks (DQN) \cite{survey}. In this work, we approach the resource allocation problem by modeling it as a decision-making process with discrete action spaces, where each action corresponds to a specific allocation strategy. The discrete nature of the problem makes DQN a well-suited choice, as it is designed to handle decision-making problems involving discrete actions by optimizing policies based on maximizing long-term rewards. 

Moreover, DQN effectively combines Q-learning with deep neural networks, allowing for efficient learning of action-value functions even in high-dimensional state spaces, which is crucial for addressing the complexity of RRM tasks. To facilitate the necessary flexibility and intelligence in future networks, the Open Radio Access Network (ORAN) Alliance has proposed a network architecture that supports a programmable, virtualized, and vendor-agnostic approach. This architecture enables the seamless integration of intelligence into various network components, offering advanced, pervasive decision-making capabilities \cite{oran1}. ORAN’s architecture is particularly crucial for managing the exponentially increasing traffic expected in 6G networks while ensuring dependable performance across all use cases. Moreover, it supports technologies like artificial intelligence (AI)-assisted RAN slicing, where different slices cater to specific 3GPP use cases such as eMBB, URLLC, and mMTC. By isolating resources into distinct slices, ORAN allows for more efficient management of service-level agreements (SLAs) and a controlled, dynamic approach to resource management.

Furthermore, ORAN facilitates the implementation of AI as a Service (AIaaS) at the edge of the network, enabling intelligent network components to assist in resource allocation and other decision-making processes. The proposed algorithm in \cite{mypaper} focuses on optimizing the allocation and utilization of resources within these network slices. Similarly, several state-of-the-art (SOTA) works leverage the intelligent aspects of ORAN, such as the near- and non-real-time RAN Intelligent Controllers (near \& non-RT RICs), to improve resource management \cite{oran2, oran3}. In addition to DRL, Explainable AI (XAI) has gained prominence as a means to enhance transparency and trust in AI systems, particularly in critical applications like network management. The integration of XAI with DRL not only improves the effectiveness of these algorithms but also ensures compliance, policy justification, and security by providing insights into the decision-making processes \cite{main, globecom1}. Recent work, including our proposed framework, combines XAI with DRL to develop explainable deep reinforcement learning (XRL) applications, or xAPPs, which are deployed on the near-RT RIC. These applications interact with various network components to enable intelligent, real-time decision-making while maintaining system trustworthiness. This paper presents the XAI framework as an xAPP capable of interfacing with multiple algorithms for different network slices. The integration of DRL-based xAPPs and rAPPs (applications running on the non-RT RIC) further enhances the scalability and adaptability of the network, ensuring efficient radio resource management across both granular and broad operational levels. The proposed work tackles challenges such as system complexity and the integration of legacy systems, using ORAN as a foundation for seamless incorporation of AI-driven solutions.

\section{Related Work}
Intra- and inter-slice resource allocation in RAN is pivotal for optimizing network efficiency across different timescales within the ORAN architecture. Current SOTA methods, as discussed in various studies \cite{Melike1, UA, UA2}, focus on maximizing network performance through algorithms that enhance xAPP collaboration and optimize aggregate KPIs. Specifically, \cite{Melike1} introduces a team learning algorithm in an ORAN setting aimed at boosting base station throughput. Similarly, research in \cite{UA, UA2} tackles the combined challenges of user association and bandwidth allocation, proposing solutions to maximize data rates in heterogeneous networks (HetNets). However, these methods often rely on single-objective reward functions, limiting their ability to address multi-faceted network demands effectively.\cite{OpenRAN} introduces an OpenRAN Gym, a comprehensive framework designed to enable AI-driven applications xApps in ORAN compliant networks. It provides detailed methodologies for utilizing xApps on real-time controllers, thereby optimizing network control and enhancing the functionality of next-generation cellular networks.

Inter-slice resource allocation studies, such as those by \cite{Melike2, twc , mypaper}, explore resource distribution among different network slices, employing techniques like correlated Q-learning to achieve higher throughput and lower queuing delays for specific slices like eMBB and URLLC. Further research, such as that presented in \cite{globecom1}, emphasizes the role of xApps in dynamic and efficient resource management across multiple RAN slices, utilizing data-driven approaches. This work showcases the potential of AI/ML technologies to adapt resource allocation dynamically to fluctuating traffic demands and operational conditions, allowing for real-time optimization of network performance. Such developments highlight the critical importance of integrating advanced AI/ML solutions in managing virtualized RANs, pointing towards a future where AI not only directs but also explains and refines resource management strategies in real-time telecommunications networks. 

The earlier discussed techniques suffer from a lack of transparency in decision-making processes, especially in dynamic and complex network environments where multiple parameters must be simultaneously optimized. The incorporation of XAI within DRL frameworks address these shortcomings, for instance, a novel approach, AI EXPLainability for the Open RAN (EXPLORA), an open source framework is highlighted in \cite{main}, 
which has been used as a baseline and extended in this research. It is an xApp developed for ORAN-compliant near-RT RIC that improves the transparency of DRL decisions in resource allocation. This advancement not only facilitates better monitoring and troubleshooting but also enhances network performance through proactive, AI-generated explanations, making AI-driven decisions more understandable and reliable. While EXPLORA represents a significant improvement, the existing procedure limit its ability to handle multiple KPIs- or quality of service (QoS)-based action steering.

These capabilities are critical for the framework to be effectively utilized across a diverse range of xAPPs, each focused on different types of network slices. Moreover, the application of XAI in \cite{main} is focused on xAPPs, whereas it could also be beneficially applied at the interslice, non-RT level on rAPPs to extend similar advantages to those DRL agents. These gaps in application and functionality are bridged with our proposal as discussed in Sections V and VI of this paper. Here, we introduce XAI framework for multi-faceted network management with aforementioned abilities. 
We detail the following contributions which are aimed at refining action steering mechanisms within these complex XAI-driven systems:
\begin{enumerate}
    \item This research proposes a refined XAI framework, XRL-QoS that enhances logical reasoning in DRL-based decision-making processes. By offering clear insights into the control actions taken by both xAPPs and rAPPs, our approach significantly improves transparency, making the evaluation of resource management actions more interpretable and actionable for operators. 
    \item Our work introduces several key enhancements that build on the existing XAI framework in \cite{main}.  While XAI in SOTA was designed to focus on optimizing single KPI, our proposed framework balances multiple KPIs dynamically. It achieves this by steering decision-making based on real-time QoS assessments across different network slices (e.g., eMBB, URLLC, mMTC). This multiobjective approach leads to more comprehensive and balanced system performance, which is crucial in the diverse operational environment of 6G networks.
    \item Our proposal XRL-QoS extends EXPLORA’s capabilities by integrating it as a standalone xAPP that operates within an ORAN-compliant near-RT RIC. This adaptation showcases the flexibility and scalability of the proposed framework in real-world environments. 
    \item A key advancement of our proposal is its applicability across multiple timescales. This feature makes the framework versatile enough to manage resources effectively in both inter-slice and intra-slice scenarios. Whether applied in offline-trained environments or near-RT online training contexts, the framework can adapt to the varying requirements of different network slices, ensuring optimal resource allocation at both granular and broader levels. This improvement demonstrates the framework’s robustness in handling both short-term fluctuations and long-term strategic planning.
\end{enumerate}

\section{ORAN based slice resource management}

\begin{figure*} [htbp]
    \centering 
    \includegraphics[width=16cm, height=7.5cm]{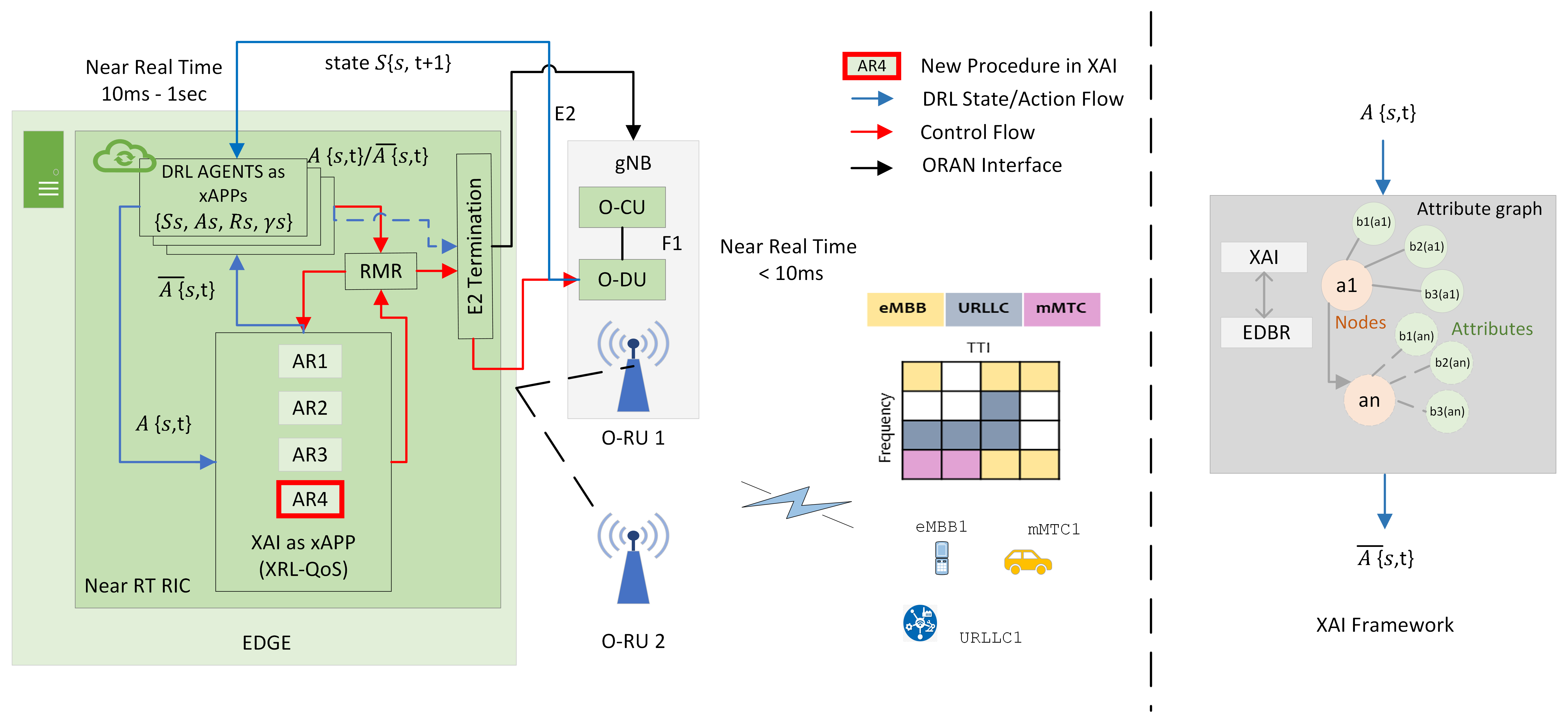}
    \caption{XAI applied DRL based Slice Resource Management in ORAN}
    \label{fig:xai}
\end{figure*}

This research paper explores an advanced architectural framework designed to enhance network slicing using a dynamic and intelligent approach. By incorporating DRL algorithms, the system aims to optimize and manage network resources dynamically across different segments of a wireless network. As shown in Figure \ref{fig:xai}, we consider ORAN-based architecture. It integrates various functional entities including the service management and orchestrator (SMO), non-RT RIC, near RT RIC, and various DRL agents deployed as rAPPs and xAPPs. It shows an XAI framework as xAPP and a RIC message router (RMR). Here, the control flow and exchanges for evaluations of the DRL agent's actions are depicted with red arrows. The red highlighted AR4 procedure is the novel XAI proposal discussed in detail in Section VI of this paper.

In the ORAN network, we study a configuration consisting of a group of users $\mathcal{K} = \{ 1, \cdots, K \}$ and ORAN radio units (ORUs) $\mathcal{M} = \{ 1, \cdots, M \}$.
Users are categorized into three slice types: eMBB, URLLC, and mMTC, represented as $\mathcal{K}_s$ where $s$ stands for each slice type, and each user set is mutually exclusive. Each ORU serves all three slices and has a set of resource blocks (RBs) $\mathcal{N} = \{ 1, \cdots, N \}$. 
RBs are organized into groups resource block groups (RBGs) following 5G NR standards with numerology $\mu = 0$. Each ORU has fixed bandwidth allocations per slice, expressed as $W_{m,s}$
Each slice, denoted by $s \in \{E,U,M\}$, is allocated a fixed number of RBGs labeled $Z_s$. An intra-slice intelligent agent is tasked with allocating RBs to end-users within these slices based on available resources. For scheduling, a threshold $\tau^{s}_{th}$ is established unique to each slice. Then the packets in the ORU buffer are prioritized as per time out reaching this threshold \cite{mypaper}. 
After scheduling, the system assesses the maximum resource utilization and its variation. This utilization metric is computed based on the total resources used by all users within a slice as defined below,
\begin{equation}
    U_{s,t} = \sum_{k \in \mathcal{K}_s} \sum_{m \in \mathcal{M}} N_{k,m,t}\ ,\  U^{max}_{s,t} =  \max_{t \in TTI} U_{s,t}
\end{equation} 
Here, $N_{k,m,t}$ is number of RBs assigned to user $k$ under ORU $m$ at time $t$. The variations of $U^{max}_{s,t}$ are observed and forwarded to inter-slice resource allocation for better reconfiguration decisions. The variation is calculated as below:
\begin{equation}
\begin{split}
    \delta_s = \frac{U^{max}_{s,t} - \psi_s}{\psi_s}  \ \ \text{where,} \ \
    \psi_s = \frac{1}{|T|} \sum_{t = 0}^{T} U^{max}_{s,t}
\end{split}
\end{equation}
The performance of each user or slice is assessed using throughput and delay metrics. The average throughput for each slice, denoted as $R^s_{avg}$, is determined by summing the throughput of all users in that slice and dividing by the total number of users in the slice. To normalize the throughput for individual users denoted as ${\bar{R}}_{k,norm}$, it is divided by their minimum QoS requirement ($R_k^{min}$). For any user $k \in \mathcal{K}_s$, the average throughput ${\bar{R}}_k$ across $T$ transmission time intervals (TTIs) is computed using Shannon's capacity formula. 
\begin{equation}
     R_{avg}^s =\frac{\sum_{k \in \mathcal{K}_s} {\bar{R}}_{k,norm}}{|\mathcal{K}_s|} =\frac{1}{|\mathcal{K}_s|}\sum_{k \in \mathcal{K}_s}\frac{{\bar{R}}_k} {R_k^{min}} 
\end{equation}
Each packet experiences delay in queuing, transmission, and processing accumulating as a total delay experienced by each packet under each user. The average delay per user is determined by calculating the mean delay across all packets. 
Further, the average delay performance per slice, $d_{avg}^s$ is calculated similar to average throughput performance calculated in equation (3) where, normalized delay ${\bar{d}}_{k,norm}$ is defined based on QoS requirements.
The exponentially increasing service demands need to be fulfilled with existing limited bandwidth. Hence, the objective of different slices is to minimize the maximum utilization of resources within each slice while achieving predefined QoS performance. This will enable better usage of limited resources without affecting QoS. It is expressed as follows: 
\begin{equation}
\begin{split}
    &\min U^{max}_{s,t} \\
    \text{subjected to:} \ \ &{P}^{QoS}_k =1 , \ k \in \mathcal{K}_s
\end{split} 
\end{equation}
\begin{equation}
    \text{Where,} \ \ {P}^{QoS}_k\ =1\ if\ \left\{{\bar{R}}_{k,norm}\geq1 \land {\bar{d}}_{k,norm}\geq1\right\}
\end{equation}

\section{DRL Agents as x/rAPP and Multi-timescale Resource Management}
The intelligent agents as xAPPs and rAPP utilize the DRL technique, specifically DQN, with unified objective functions at the intra-slice level aimed at minimizing maximum resource utilization. Conversely, inter-slice agents reconfigure the fixed RBG pool across different slices based on performance KPIs and resource usage data from intra-slice agents. Distinct intelligent agents are tailored for eMBB, URLLC, and mMTC slices. 
Each DQN agent establishes a different threshold $\tau_{th}^s$  
for each slice which is crucial for scheduling as explained in above section and described in \cite{mypaper}. 
The Markov decision process (MDP) for the intelligent agents, comprises a tuple \(\left\{S_s, A_s, R_s, \Gamma_s\right\}\), representing state, action, reward, and discount factor respectively. The MDP for both intra- and inter-slice DRL agents, associated with xAPP and rAPP are given below described thoroughly in Sections V-A and V-B of \cite{mypaper}. The state space for intra-slice DRL agents is defined as channel matrix and number of bits in the buffer. Whereas, the action space is discretized timeout threshold values $\tau^{s}_{th}$. The set of $\text{D}^{s}$ discrete values ranging from $\tau_{\text{min}}$ to \(\tau_{\text{max}}\) hence the action space and reward can be expressed as below:
\begin{equation}
    A_s = \tau^{s}_{th} \in \{\tau^{s}_{\text{min}}, \tau^{s}_{\text{min}} + \tau^{s}_{step}, \ldots, \tau^{s}_{\text{max}} - \tau^{s}_{step}, \tau^{s}_{\text{max}}\}
\end{equation}
\begin{equation}
        {R}^i_s\left(S_s,A_s\right) = \alpha_s \cdot U^{max}_{s,i}+ \beta_s \cdot R_{avg}^{s,i}
\end{equation}
The MDP for inter-slice intelligent agent using DQN is defined as shown below:
\begin{equation}
    S =\left\{R_{avg}^{s}, U^{max}_{s}, \delta_s\right\} \ \text{for}\ s\in \{ E, U, M\}
\end{equation}
The action space for this intelligent agent is defined as the combination of all available RBGs $Z^{comb}$ distributed among slices $s$. The action space (has $D$ combinations) and the reward are defined as below: 
\begin{equation}
    A =  \{Z^{comb}_1, \ldots , Z^{comb}_D\}
\end{equation}
\begin{equation}
        {R}^i\left(S,A\right) = \sum_{s} \left( R_{avg}^{s, i} -  d_{avg}^{s, i} \right) %
\end{equation}
The calculations for the Q value, target Q value, and loss function are detailed in \cite{book2}. The values for learning rate $\Omega_s$, as well as other hyper-parameters are indicated in the Table \ref{tab:Param}. 

\section{XRL: EXPLORA framework}
EXPLORA is an open source AI/ML explainability framework for ORAN which has been made available as an xAPP for ORAN-compliant near-RT RICs, facilitating network-oriented explanations and evaluation of DRL agents control actions \cite{main}. This framework is designed to enhance explainability in DRL systems used within the ORAN architecture. It uses an attributed graph model to link DRL agent actions to input states, helping to clarify the AI's decision-making process in network environments. Figure \ref{fig:xai} shows XAI framework that demonstrate these attribute graph showcasing nodes \{$a_{t}, a_n$\} and attributes \{$(b_1(a_{1}), b_n(a_{1}, b_n(a_{n}),....$\}. 

This makes the behavior of DRL models more transparent, aiding in their deployment and management in ORAN systems, especially for tasks like resource allocation. The transparency is achieved by synthesizing the network-oriented explanation. In our framework the XAI module is connected to both DRL rAPP and DRL xAPP, as well as the data base and SMO. It ensures that all machine learning-driven decisions are transparent, providing insights back to the DRL agents and influencing slice management policies. It accesses historical data from the database to aid in decision-making and policy refinement. There are three default strategies/procedures available to drive the transparency namely, AR1:"Max-Reward" , AR2:"Min-reward", AR3:"Improved bit-rate".

\section{Proposed XRL: QoS-based action steering}
Our study presents a detailed examination of an intelligent network slicing architecture empowered by DRL agents. We link these agents to an eXplainability module, XAI as xAPP, XRL-QoS, as shown in Figure \ref{fig:xai}.  
We adhere to the algorithmic flow of EXPLORA outlined in Section V, Algorithm 1 of the research work \cite{main}, incorporating the new proposed procedure AR4 as shown below and included in XAI module with red highlighted box in Figure \ref{fig:xai}. For the extended XAI framework i.e. given new procedure the attribute graph can be expressed with nodes {$a_{br}, a_d$} and attributes {$b_j(a_{br}), b_k(a_{d})$}.
\begin{algorithm}
\caption{New Procedure for QoS-based action steering}
\begin{algorithmic}[1]
\Procedure {AR\_4}{$Q, a_t$}
\State $j \leftarrow \text{Index of } R^{s,i} \text{ KPI in the attributes } b(\cdot) \in G$
\State $k \leftarrow \text{Index of } d^{s,i} \text{ KPI in the attributes } b(\cdot) \in G$
\State $a_{br}=\arg \max_{a} \{b_j(a) : b_j(a) \in b(a) \in (w, b(a)) \in Q\}$
\State $a_{d} = \arg \max_{a} \{b_k(a) : b_k(a) \in b(a) \in (w, b(a)) \in Q\}$
\If {$b_j(a_{br}) > b_j(a_t) \And b_k(a_{d}) > b_k(a_t)$}
\State $a_t \leftarrow \arg \max_{a_t} a_{br}, a_d $
\EndIf
\EndProcedure
\end{algorithmic}
\end{algorithm}

In this study, we modify the approach used in AR1, AR2, and AR3 where a single KPI was utilized to map the argmax. Instead, we concurrently use two KPIs: throughput \(R^{s,i}\) and delay \(d^{s,i}\) as defined in Section IV of this paper. We introduce a new procedure termed “QoS based action steering”. Unlike other three existing procedures, AR4 replaces an action \(a_t\) calculated by the DRL agent (anticipated to yield a low KPIs or reward) with another action \(a_G\) expected to deliver high throughput and low delay. This can be achieved by extracting the attributes $b(a_t)$ and $b(a_G)$ from the attributed graph $G$, and computing the expected reward using the KPI values. This strategy aims to equitably balance the actions across all slices by distributing the importance of KPIs. Thus, regardless of whether a DRL agent or xAPP is primarily focused on an eMBB, URLLC, or mMTC slice, it considers all crucial KPIs, ensuring a balanced logical explanation for action selections.

\begin{figure} [htbp]
    \centering 
    \includegraphics[width=\linewidth]{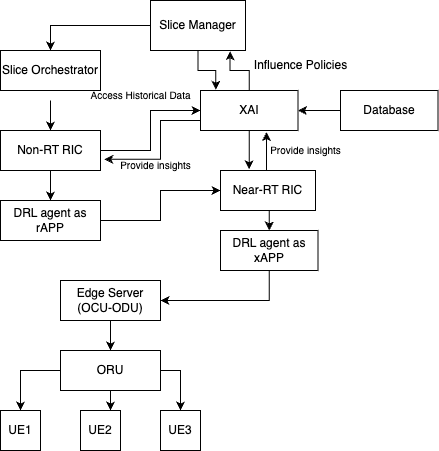}
    \caption{XAI and DRL based policies for resource allocation}
    \label{fig:adapt}
\end{figure}

Figure \ref{fig:adapt} shows policy and information flow while applying XAI as xAPP to benefit it with other xAPPs and rAPPs. The flow and interconnections between different entities highlight a comprehensive approach to real-time and non-real-time network management, crucial for next-generation wireless networks. Here, the DRL rAPP interacts with the near-RT RIC to translate high-level policies from the non-RT RIC into actionable, real-time decisions, thereby facilitating immediate adjustments in network behavior. DRL xAPP works in conjunction with the near-RT RIC but focuses specifically on optimizing edge-based processes. It sends commands to the edge server, which directly interfaces with the operational units of the network, including OCU, ODU, ORU. These components manage the connectivity and communication protocols with end-user devices. They are crucial for implementing the real-time decisions made by the DRL xAPP at the edge of the network.

\section{Simulation Setup}
Table \ref{tab:Param} shows QoS requirement, traffic model, DRL agent settings, etc. The simulator is built on python based environment using open-AI-gym, Keras, TensorFlow etc. modules. 
Refer to Section VI and VII of \cite{mypaper} for more details. We utilize the same configuration for a 5G NR environment with $\mu=0$, including a sub-carrier spacing of 15 kHz across 12 sub-carriers, resulting in a bandwidth of 180 kHz for each RB, and each RBG comprising 6 RBs. We allocate a total of 14 RBGs in three network slices. The simulation parameters for transmission power, antenna placement, noise figures, variance values, and path loss exponent follow the ITU-R recommendations (tables A1-2)\cite{itu1} and adhere to the ORAN standards. The network setup includes 3 ORUs distributed across the three slices, supporting a total of 9 users, with 3 users per slice. 

The intra-slice DQN agents operate every 10 TTIs and inter-slice agents every 200 TTIs. Furthermore, the open-source repository of EXPLORA extened to XRL-QoS, which functions as a separate xAPP, is adapted for use within our own simulator. As mentioned previously, we have expanded this framework by incorporating a new procedure. This xAPP assesses the actions taken by each DRL agent to provide network-oriented explanations. The configuration of XAI and existing x/rAPPs is illustrated in Figure \ref{fig:xai}.

\begin{table}[htbp]
\centering
\caption{Simulation Parameters}
\label{tab:Param}
\begin{tabularx}{\linewidth}{X X}
\toprule
Parameter & Value  \\
\midrule
 $\bm{R}_k^{\text{min}}$ (Mbps)& E=16, U=3.8, M=0.5 \\
$\bm{d}_k^{\text{max}}$ (ms)& E=10, U=2, M=20  \\
Periodic Deterministic PAR (ms) & E=0.5, U=1, M=0.5 \\
Packet Size (Bytes) & E=1024,U=480,M=32 \\
xAPP DRL N/W Arch & 64 $\times$ 256 $\times$ 256 \\ 
xAPP DRL $\Omega$ & $1 \times 10^{-4}$ \\ 
xAPP DRL Batch size & 64 \\ 
rAPP DRL N/W Arch & 256 $\times$ 256 \\ 
rAPP DRL $\Omega$ & $1 \times 10^{-4}$ \\ 
rAPP DRL Batch size & 64 \\ 
TN Update & 100 \\ 
Activation \& Optimizer & ReLU \& Adam \\ 
\bottomrule
\end{tabularx}
\end{table}
\section{Results}
\begin{figure}[htbp]
    \centering
    \begin{subfigure}{0.46\textwidth}
        \centering
        \includegraphics[width=\textwidth]{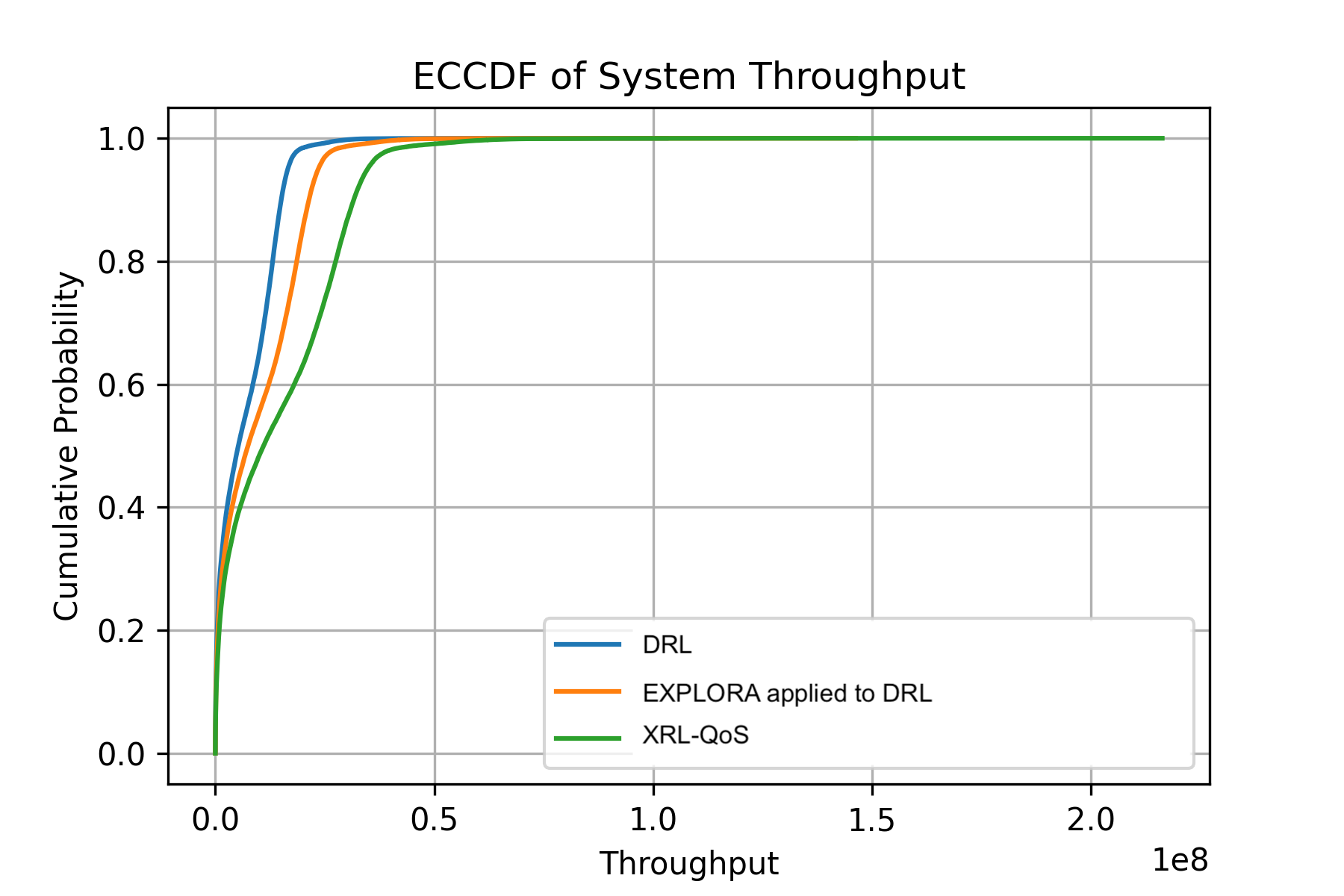}
        \caption{System Throughput ($R_{avg}$) Performance Comparison} %
        \label{subfig:subfigure1}
    \end{subfigure}
    \hfill
    \begin{subfigure}{0.46\textwidth}
        \centering
        \includegraphics[width=\textwidth]{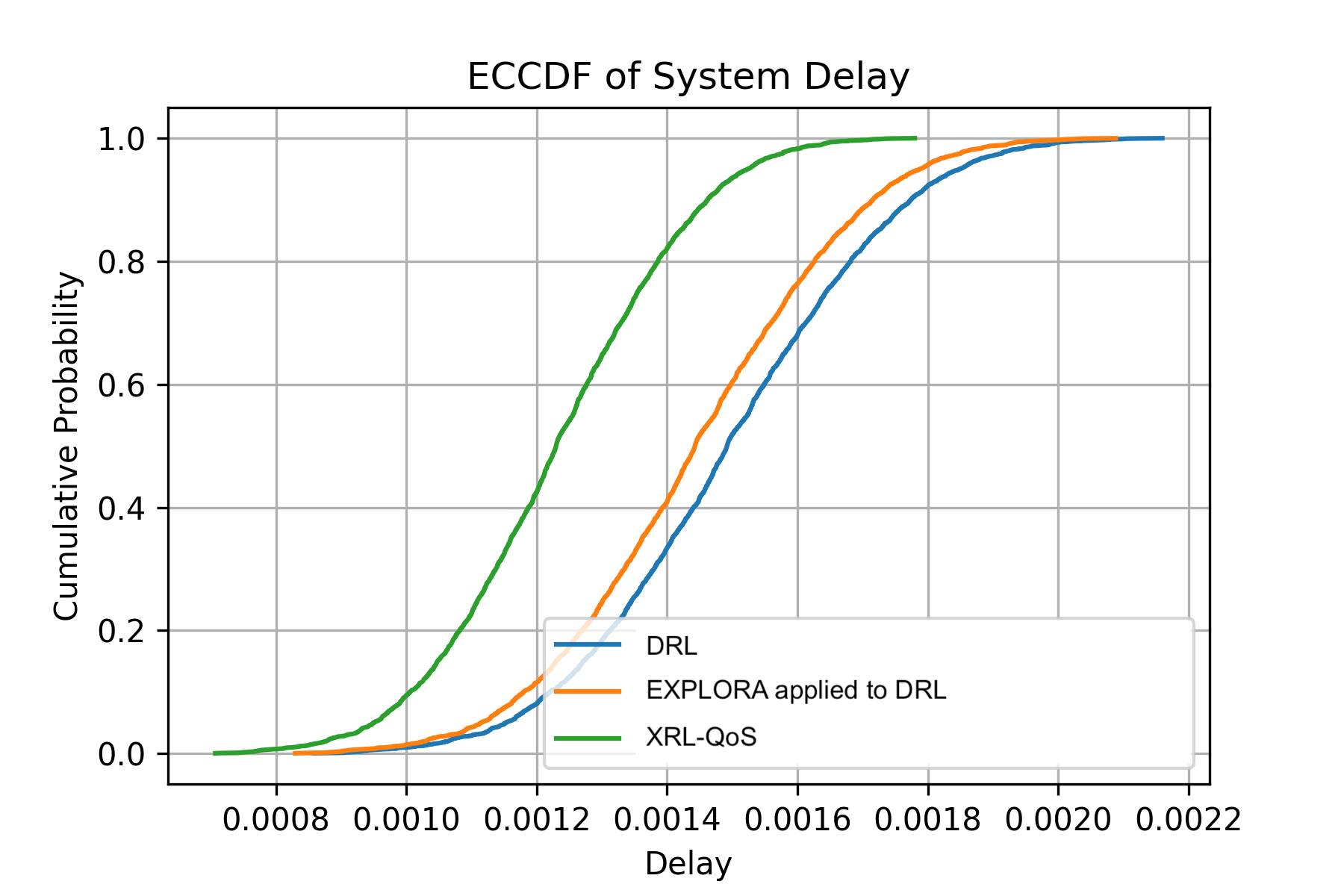}
        \caption{System Delay ($d_{avg}$) Performance Comparison}
        \label{subfig:subfigure2}
    \end{subfigure}
    \caption{System KPIs}
    \label{fig:R1}
\end{figure}

The results are obtained using the setup previously discussed. Intra-slice intelligent agents typically achieve convergence within 180–250 DQN runs. Additionally, the inter-slice intelligent agent, trained offline for 100 seconds, reaches convergence in the reward function within the first 50 runs \cite{mypaper}. Subfigures \ref{subfig:subfigure1} and \ref{subfig:subfigure2} display the system KPI performance comparisons for throughput and delay, respectively. The Empirical Cumulative Distribution Function (ECCDF) is used to compare the performance among the DRL-based agents \cite{mypaper}, the EXPLORA-applied DRL agents \cite{main}, and the proposed work i.e. XRL-QoS agents. The DRL employs a similar simulation setup as previously described, without any additional evaluation on top of the DRL control actions. In contrast, the EXPLORA-applied DRL utilizes the default EXPLORA settings over the DRL, while XRL-QoS incorporates a new procedure along with the default EXPLORA settings. Subfigure \ref{subfig:subfigure1} highlights the system throughput performance across all users within the three slices. 

It is evident that throughput increases when the XAI framework is used with DRL agents. The XRL-QoS provides highest throughput compared to the other methods and enhances overall system performance. Subfigure \ref{subfig:subfigure2} demonstrates that the minimum system delay is reduced by 5–12\% in each slice due to the proposed intra-slice agent's separate timeout thresholds, which facilitate more informed decisions through QoS-based, network-oriented explanations. Existing procedures in EXPLORA, such as AR3, focus on maximizing action selection with high transmit bits; however, the newly proposed procedure balances delay and throughput KPIs, significantly improving performance for delay compared to the default EXPLORA method. Furthermore, the XRL-QoS algorithm reduces the maximum resource utilization by 12\% compared to the DRL method and by 5.2\% compared to the EXPLORA-applied DRL method, facilitating more effective resource reconfiguration at the inter-slice level.

\section{Conclusion and Future Scope}
Our contributions collectively improve the strategic implementation of DRL methods and exemplify the complex yet impactful nature of AI-driven networks pivotal for 6G. 
The results provide insights on improvements in performance of DRL-based xAPP and rAPP for ORAN architecture. 
The framework improves the monitoring, proactive explanation for control actions and its steering in existing DRL techniques. The proposed XRL-QoS improves the system throughput performance by 7.65\% compared to the DRL and 4.82\% to EXPLORA applied DRL as given in \cite{mypaper} and \cite{main} respectively. Whereas, it improves delay performance by almost 14\% compared to DRL method. The overall system performance is elevated in addition to improvement in slice targeted KPIs. Our work verifies the adaptability of EXPLORA to our own test set up as well as its openness to include different procedures. This work validates application of XAI and its benefits at different timescales for intra- and inter-slice level resource management. This could be further explored to improve the performance of various other xAPPs. 

\section*{Acknowledgment}
The above work is conducted as a part of MSCA ITN SEMANTIC Project (861165).



\begin{thebibliography}{1}

\bibitem{survey}
N. Luong, D. Hoang, S. Gong, et al, “Applications of Deep Reinforcement Learning in Communications and Networking: A Survey", \emph{IEEE Communications Surveys \& Tutorials}, vol. 22, no. 4, pp. 2664-2732, 2020. DOI: 10.1109/COMST.2020.2988348

\bibitem{oran1} 
ORAN Specifications, “ORAN Architecture Specification", Version 1.0.0, 8 July 2019.


\bibitem{mypaper}
S. Mhatre, F. Adelantado, K. Ramantas and C. Verikoukis, "AIaaS for ORAN-based 6G Networks: Multi-time Scale Slice Resource Management with DRL," ICC 2024 - IEEE International Conference on Communications, Denver, CO, USA, 2024, pp. 5407-5412, doi: 10.1109/ICC51166.2024.10622601.

\bibitem{oran2}
ORAN Technical Specifications, “ORAN WG1 Slicing-Architecture-v07.00: Technical Specification", July 2022.
\bibitem{oran3}
ORAN TR, “ORAN.WG1.Use-Cases-Analysis-Report-v08.00", 2022.

\bibitem{main}
C. Fiandrino, L. Bonati, S. D'Oro, M. Polese, T. Melodia, and J. Widmer, 2023, “EXPLORA: AI/ML EXPLainability for the Open RAN," Proc. ACM Netw. 1, CoNEXT3, Article 19 (December 2023), 26 pages. https://doi.org/10.1145/3629141.

\bibitem{globecom1}
L. Bonati, M. Polese, S. D'Oro, S. Basagni and T. Melodia, “Intelligent Closed-loop RAN Control with xApps in OpenRAN Gym," European Wireless 2022; 27th European Wireless Conference, Dresden, Germany, 2022, pp. 1-6.

\bibitem{Melike1}
H. Zhang, H. Zhou and M. Erol-Kantarci, “Team Learning-Based Resource Allocation for Open Radio Access Network (O-RAN)", \emph{IEEE International Conference on Communications}, Seoul, Korea, 2022, pp. 4938-4943, doi: 10.1109/ICC45855.2022.9838763.

\bibitem{UA}
Z. Li, M. Chen, K. Wang, C. Pan, N. Huang and Y. Hu, “Parallel Deep Reinforcement Learning Based Online User Association Optimization in Heterogeneous Networks", \emph{IEEE International Conference on Communications Workshops (ICC Workshops)}, Dublin, Ireland, 2020, pp. 1-6, doi: 10.1109/ICCWorkshops49005.2020.9145209.

\bibitem{UA2}
Q. Zhang, Y. -C. Liang and H. V. Poor, “Intelligent User Association for Symbiotic Radio Networks Using Deep Reinforcement Learning", \emph{IEEE Transactions on Wireless Communications}, vol. 19, no. 7, pp. 4535-4548, July 2020, doi: 10.1109/TWC.2020.2984758.

\bibitem{OpenRAN}
L. Bonati, M. Polese, S. D’Oro, S. Basagni and T. Melodia, “OpenRAN Gym: An Open Toolbox for Data Collection and Experimentation with AI in O-RAN," 2022 IEEE Wireless Communications and Networking Conference (WCNC), Austin, TX, USA, 2022, pp. 518-523, doi: 10.1109/WCNC51071.2022.9771908. 

\bibitem{Melike2}
H. Zhou, M. Elsayed and M. Erol-Kantarci, “RAN Resource Slicing in 5G Using Multi-Agent Correlated Q-Learning", \emph{IEEE 32nd Annual International Symposium on Personal, Indoor and Mobile Radio Communications (PIMRC)}, Helsinki, Finland, 2021, pp. 1179-1184, doi: 10.1109/PIMRC50174.2021.9569358.

\bibitem{twc}
C. V. Nahum et al., “Intent-aware Radio Resource Scheduling in a RAN Slicing Scenario using Reinforcement Learning", \emph{IEEE Transactions on Wireless Communications}, August 2023,  doi: 10.1109/TWC.2023.3297014.


\bibitem{book2}
S. Sutton and A. G. Barto, “Reinforcement Learning: An Introduction", MIT Press, 2018
\bibitem{itu1}
ITU-R M.2412-0, “Guidelines for evaluation of radio interface technologies for IMT-2020", 2020


\end{thebibliography}
\end{document}